\documentstyle[12pt]{article}
\begin{document}

\begin{titlepage}

\centerline{\LARGE
Self-Adjointness of the Dirac Hamiltonian}
\vspace{.3cm}
\centerline{\LARGE
and Fermion Number Fractionization in}
\vspace{.3cm}
\centerline{\LARGE
the Background of a Singular Magnetic Vortex}
\vspace{1cm}
\begin{center}
{\large Yu.A.~Sitenko}\\
\mbox{} \\
{\it Bogolyubov Institute for Theoretical Physics, National Academy of Sciences of Ukraine,} \\
{\it  252143 Kiev, Ukraine}
\end{center}

\vskip 1 cm
\begin{abstract}
The method of self-adjoint extensions is employed to determine the
vacuum quantum numbers induced by a singular static magnetic vortex in
$2+1$-dimensional spinor electrodynamics. The results obtained are
gauge-invariant and, for certain values of the extension parameter,
both periodic in the value of the vortex flux and possessing definite parity
with respect to the charge conjugation.
\end{abstract}

\medskip

\ \ \ \ \ PACS numbers: 03.65.Bz, 12.20.Ds, 11.30.Er, 11.30.Qc
\end{titlepage}

\setcounter{page}{1}

Twenty years ago Jackiw and Rebbi [1] have discovered that
the fermionic vacuum under certain conditions can acquire rather
unusual properties, with the vacuum fermion number becoming nonzero
and even noninteger. This phenomenon named as the fermion number
fractionization is appearing in a variety of quantum field systems of
different spatial dimensionalities (see the reviews in
Refs.[2,3]. Some aspects of the fermion number
fractionization in spaces with nontrivial topology will be elucidated
in the present Letter.

In the framework of the secondly quantized theory the vacuum value of
the fermion number operator is related to the spectral asymmetry of
the Dirac Hamiltonian (see e.g. Ref.[3]),
\begin{equation}
\langle N\rangle=-{1\over2}{\sum\!\!\!\!\!\!\!\int\limits_E}{\rm sgn}(E),
\end{equation}
where ${\rm sgn}(u)=\left\{\begin{array}{rl}1,&u>0\\-1,&u<0\end{array}\right.$ is the
sign function and the symbol ${\displaystyle {\sum\!\!\!\!\!\!\!\int\limits_E}}$ implies the
summation over the discrete and the integration (with a definite
measure) over the continuous parts of the energy spectrum,
$H\psi(\vec{x})=E\psi(\vec{x})$, the operator $H$ considered in this
Letter is the Dirac Hamiltonian in the background of an external
static vector field:
\begin{equation}
H=-i\vec{\alpha}[\vec{\partial}-i\vec{V}(\vec{x})]+\beta m;
\end{equation}
to regularize the sum(integral) in Eq.(1) at $|E|\to\infty$, the
factor $(E^2)^{-s}$ or $\exp(-tE^2)$ can be inserted, with the limit
$s\to0_+$ or $t\to0_+$ taken after implementing all
summations(integtations). In a flat two-dimensional space
$\bigl(\vec{x}=(x^1,x^2)\bigr)$ the vacuum fermion number is
calculated to be [4]
\begin{equation}
\langle N\rangle=-{1\over2}{\rm sgn}(m)\Phi,
\end{equation}
where $\Phi={1\over2\pi}\int d^2x B(\vec{x})$ is the total flux (in
the units of $2\pi$) of the external magnetic field strength
$B(\vec{x})=\vec{\partial}\times \vec{V}(\vec{x})$ piercing the
two-dimensional space (plane); note that the mass parameter $m$ in
Eq.(2) can take both positive and negative values in two and any even
number of spatial dimensions.

It should be emphasized, however, that Eq.(3) is valid for  regular
external field configurations only, i.e. $B(\vec{x})= B_{\rm
reg}(\vec{x})$, where $B_{\rm reg}(\vec{x})$  is a piece-wise
continuous function that can grow at most as
$O\bigl(|\vec{x}-\vec{x}_s|^{-2+\varepsilon}\bigr)$ $(\varepsilon>0)$
at separate points; as to a vector potential
$\vec{V}(\vec{x})=\bigl(V_1(\vec{x}),V_2(\vec{x})\bigr)$, it is
unambiguously defined everywhere on the plane. The regular
configuration of an external field polarizes the vacuum locally, and
Eq.(3) is just the integrated version of the linear relation between
the vacuum fermion number density and the magnetic field strength.

One can ask the following question: whether the nonlocal effects of
the external field background are possible, i.e., if the spatial
region of nonvanishing field strength is excluded, whether there will
be vacuum polarization in the remaining part of space? For the
positive answer it is necessary, although not sufficient, that the
latter spatial region be of nontrivial topology [5] (see
also Ref.[6]). However, the condition on the boundary of
the excluded region has not been completely specified. In the present
Letter this point will be clarified by considering the whole set of
boundary conditions which are compatible with the self-adjointness
of the Dirac Hamiltonian in the remaining region.

We shall be interested in the situation when the volume of the
excluded region is shrinked to zero, while the global characteristics
of the external field in the excluded region is retained nonvanishing.
This implies that singular, as well as regular, configurations of
external fields have to be considered. In particular, in two spatial
dimensions the magnetic field strength is taken to be a distribution
(generalized function)
\begin{equation}
B(\vec{x})=B_{\rm reg}(\vec{x})+2\pi\Phi^{(0)}\delta(\vec{x}),
\end{equation}
where $\Phi^{(0)}$ is the total magnetic flux (in the units of $2\pi$)
in the excluded region which is placed at the origin $\vec{x}=0$. As
to the vector potential, it is unambiguously defined everywhere with
the exception of the origin, i.e. the limiting value
$\lim_{|\vec{x}|\to0}\vec{V}(\vec{x})$ does not exist, or, to be more
precise, a singular magnetic vortex is located at the origin
\begin{equation}
\lim_{|\vec{x}|\to0}\vec{x}\times\vec{V}(\vec{x})=\Phi^{(0)}.
\end{equation}
Certainly, a plane has trivial topology, $\pi_1=0$, while a plane with
a puncture where the vortex is located has nontrivial topology,
$\pi_1={{\rm Z}}$; here ${{\rm Z}}$ is the set of integer numbers and
$\pi_1$ is the first homotopy group of the surface.

The total magnetic flux through the punctured plane is obviously
defined as
\begin{equation}
\Phi={1\over2\pi}\int d^2x B_{\rm
reg}(\vec{x})={1\over2\pi}\int\limits_0^{2\pi}d\varphi
\bigl[\vec{x}\times\vec{V}(\vec{x})\bigr]\biggm|^{r=\infty}_{r=0},
\end{equation}
where the polar coordinates $r=|\vec{x}|$ and
$\varphi=\arctan(x^2/x^1)$ are introduced.

The Dirac equation with the Hamiltonian (2) on a punctured plane is
invariant with respect to the gauge transformations
\begin{equation}
G: \vec{V}(\vec{x})\to\vec{V}(\vec{x})+\vec{\partial}\Lambda(\vec{x}),
\quad\psi(\vec{x})\to{\rm e}^{i\Lambda(\vec{x})}\psi(\vec{x}).
\end{equation}
Although the vector potential in any gauge is single-valued on a
punctured plane, this is not the case for the gauge function
$\Lambda(\vec{x})$. Since the magnetic flux $\Phi$(6) (and the field
strength $B_{\rm reg}(\vec{x})$) remains invariant under gauge
transformations, the most general condition on  $\Lambda(\vec{x})$
takes the form
\begin{equation}
\Lambda(r,\varphi+2\pi)=\Lambda(r,\varphi)+2\pi\Upsilon_\Lambda,
\end{equation}
where $\Upsilon_\Lambda$ is the independent of $r$ and $\varphi$
parameter of the gauge transformation; incidentally the magnetic flux
of the vortex $\Phi^{(0)}$(5) is changed: $\Phi^{(0)}\to\Phi^{(0)}+
\Upsilon_\Lambda$. If one takes a single-valued wave function,
$\psi(r,\varphi+2\pi)=\psi(r,\varphi)$, then, after applying a gauge
transformation to it, one gets a wave function satisfying the condition
$({\rm e}^{i\Lambda}\psi)(r,\varphi+2\pi)={\rm
e}^{i2\pi\Upsilon_\Lambda}({\rm e}^{i\Lambda}\psi)(r,\varphi)$. Thus
the set of wave functions on a punctured plane is much richer than
that of wave functions on a plane without a puncture (in the latter
case only the gauge transformations with $\Upsilon_\Lambda=0$ are
admissible). Certainly, there are no reasons to impose the condition
of single-valuedness on the initial function, and in the most general
case one takes
\begin{equation}
\psi(r,\varphi+2\pi)={\rm
e}^{i2\pi\Upsilon}\psi(r,\varphi), \end{equation}
and after applying a gauge transformation one gets \begin{equation}
({\rm e}^{i\Lambda}\psi)(r,\varphi+2\pi)={\rm
e}^{i2\pi(\Upsilon+\Upsilon_\Lambda)}({\rm
e}^{i\Lambda}\psi)(r,\varphi).
\end{equation}
Therefore, if one admits singular gauge transformations
$(\Upsilon_\Lambda\neq0)$, as well as regular ones
$(\Upsilon_\Lambda=0)$, then one has to consider wave functions
defined on a plane with a cut which starts from the puncture and goes
to infinity in the radial direction at, say, the angle
$\varphi=\varphi_c$. The boundary conditions on the sides of the cut
are globally parametrized by the values of $\Upsilon$.

All this can be presented in a more refined way, using the notion of a
self-adjoint extension of a Hermitian (symmetric) operator. The
orbital angular momentum operator, $-i\partial_\varphi$, entering the
Dirac Hamiltonian(2) is Hermitian, but not self-adjoint, when defined
on the domain of functions satisfying, say,
$\psi(r,\varphi_c+2\pi)=\psi(r,\varphi_c)=0$; this operator has the
deficiency index equal to (1,1). The use of the Weyl--von Neumann
theory of self-adjoint extension [7] yields that
$-i\partial_\varphi$ becomes self-adjoint, when defined on the domain
of functions satisfying Eq.(9) with $\varphi=\varphi_c$, where the
values of $\Upsilon$ parametrize the family of extensions. It should
be stressed that    $\Upsilon$, as well as $\Phi^{(0)}$, is changed
under the singular gauge transformations (compare Eqs. (9) and (10)),
while the difference  $\Phi^{(0)}-\Upsilon$ remains invariant.

Let us turn now to the boundary condition at the puncture $\vec{x}=0$.
In the following our concern will be in the case when the regular part
of the magnetic field is absent, $B_{\rm reg}(\vec{x})=0$. Then, in
the representation with $\alpha_1=\sigma_1$,  $\alpha_2=\sigma_2$, and
$\beta=\sigma_3$ ($\sigma_j$ are the Pauli matrices) the spinor wave
function satisfying the Dirac equation and the condition (9) has the
form
\begin{equation}
\psi(\vec{x})=\sum_{n\in{\rm Z}}\left(\begin{array}{l}f_n(r)\exp[i(n+\Upsilon)
\varphi]\\g_n(r)\exp[i(n+1+
\Upsilon)\varphi]\end{array}\right),
\end{equation}
where the radial functions, in general, are
\begin{equation}
\left(\begin{array}{l}f_n(r)\\g_n(r)\end{array}\right)=\left(\begin{array}{c}C^{(1)}_n(E)J_{n-\Phi^{(0)}+
\Upsilon}(kr)+C^{(2)}_n(E)Y_{n-\Phi^{(0)}+\Upsilon}(kr)\\
{ik\over E+m}\bigl[C^{(1)}_n(E)J_{n+1-\Phi^{(0)}+
\Upsilon}(kr)+C^{(2)}_n(E)Y_{n+1-\Phi^{(0)}+\Upsilon}(kr)\bigr]\end{array}\right),
\end{equation}
$k=\sqrt{E^2-m^2}$, $J_\mu(z)$ and $Y_\mu(z)$ are the Bessel and the
Neumann functions of the order $\mu$. It is clear that the condition
of regularity at $r=0$ can be imposed on both $f_n$ and $g_n$ for all
$n$ in the case of integer values of the quantity
$\Phi^{(0)}-\Upsilon$ only. Otherwise, the condition of regularity at
$r=0$ can be imposed on both $f_n$ and $g_n$ for all but $n=n_0$,
where
\begin{equation}
n_0={[\![}\Phi^{(0)}-\Upsilon{]\!]},
\end{equation}
${[\![} u{]\!]}$ is the integer part of the quantity $u$ (i.e. the
integer which is less than or equal to $u$); in this case at least one
of the functions, $f_{n_0}$ or $g_{n_0}$, remains irregular, although
square integrable, with the asymptotics $r^{-p}$ $(p<1)$ at $r\to0$
[8].  The question arises then, what boundary condition, instead of
regularity, is to be imposed on $f_{n_0}$ and $g_{n_0}$ at $r=0$ in
the latter case?

To answer this question, one has to find the self-adjoint extension
for the partial Hamiltonian  corresponding to the mode with $n=n_0$.
If this Hamiltonian is defined on the domain of regular at $r=0$
functions, then it is Hermitian, but not self-adjoint, having the
deficiency index equal to (1,1). Hence the family of self-adjoint
extensions is labeled by one real continuous parameter denoted in the
following by $\Theta$. It can be shown (see Ref.[9]) that,
for the partial Hamiltonian to be self-adjoint, it has to be defined
on the domain of functions satisfying the boundary condition \begin{equation}
\lim_{r\to0}\cos\biggl({\Theta\over2}+{\pi\over4}\biggr)\biggl
(|m|r\biggr)^Ff_{n_0}(r)=i\lim_{r\to0}\sin\biggl({\Theta\over2}+{\pi\over4}
\biggr)\biggl(|m|r\biggr)^{1-F}g_{n_0}(r),
\end{equation}
where
\begin{equation}
F={\{\hspace{-3.3pt}|}\Phi^{(0)}-\Upsilon{\}\!\!\!|},
\end{equation}
${\{\hspace{-3.3pt}|} u{\}\!\!\!|}$ is the fractional part of the quantity $u$,
${\{\hspace{-3.3pt}|} u{\}\!\!\!|}=u-{[\![} u{]\!]}$, $0\leq{\{\hspace{-3.3pt}|} u{\}\!\!\!|}<1$; note here that
Eq.(14) implies that $0<F<1$, since in the case of $F=0$ both
$f_{n_0}$ and $g_{n_0}$ satisfy the condition of regularity at $r=0$.

Using the explicit form of the solution to the Dirac equation in the
background of a singular magnetic vortex, it is straightforward to
calculate the vacuum fermion number (1) induced on a punctured plane.
As follows already from the preceding discussion, the vacuum fermion
number vanishes in the case of integer values of $\Phi^{(0)}-\Upsilon$
$(F=0)$, since this case is indistinguishable from the case of trivial
background, $\Phi^{(0)}=\Upsilon=0$. In the case of noninteger values
of $\Phi^{(0)}-\Upsilon$ we get (details will be published elsewhere)
\begin{eqnarray}
&&\langle N\rangle=-{1\over2}
{\rm sgn}(m)\biggl(F-{1\over2}\biggr)-{1\over4\pi}\int\limits_1^\infty
{dv\over v\sqrt{v-1}}{{\rm
sgn}(m)(Av^F-A^{-1}v^{1-F})+4\bigl(F-{1\over2}\bigr)\bigl(v-1\bigr)\over
Av^F+2{\rm sgn}(m)+A^{-1}v^{1-F}},\nonumber\\&&\ \ \ \ \ \ \ \ 0<F<1,
\end{eqnarray}
where
\begin{equation}
A=2^{1-2F}{\Gamma(1-F)\over\Gamma(F)}\tan
\biggl({\Theta\over2}+{\pi\over4}\biggr),
\end{equation}
$\Gamma(u)$ is the Euler gamma-function. At half-integer values of
$\Phi^{(0)}-\Upsilon$ Eqs.(16) and (17) yield
\begin{equation}
\langle N\rangle=-{1\over\pi}\arctan\biggl\{\tan\biggl[{\Theta\over2}+
{\pi\over4}\bigl(1-{\rm sgn}(m)\bigr)\biggr]\biggr\}\quad
\left(F={1\over2}\right);
\end{equation}
note that the latter relation in the case of $m>0$ was obtained
earlier in Ref.[10]. We get also the relations
\begin{equation}
\lim_{F\to0}\langle N\rangle={1\over2}{\rm sgn}\biggl[{\rm sgn}(m)+
A\bigm|_{F=0}\biggr]
\end{equation}
and
\begin{equation}
\lim_{F\to1}\langle N\rangle=-{1\over2}{\rm sgn}\biggl[{\rm sgn}(m)+
A^{-1}\bigm|_{F=1}\biggr],
\end{equation}
indicating that the vacuum fermion number is not, in general,
continuous at integer values of $\Phi^{(0)}-\Upsilon$; the limiting
values (19) and (20) differ from the value at $F=0$ exactly, the latter
being equal, as noted before, to zero.

It is obvious that the vacuum fermion number at fixed values of
$\Upsilon$   and $\Theta$ is periodic in the value of
$\Phi^{(0)}$. This feature (periodicity in $\Phi^{(0)}$) is also
shared by the quantum-mechanical scattering of a nonrelativistic
particle in the background of a singular magnetic vortex, known as the
Aharonov-Bohm effect [11]. Since there appear assertions in
the literature which deny the periodicity of the vacuum fermion number
in $\Phi^{(0)}$ [12,13], the following comments on the
result (16) will be clarifying.

Under the charge conjugation,
\begin{equation}
C:\quad \vec{V}\to-\vec{V},\quad \psi\to\sigma_1\psi^*,\quad
\Upsilon\to-\Upsilon,
\end{equation}
the fermion number operator and its vacuum value are to be odd, $N\to
-N$ and $\langle N\rangle\to -\langle N\rangle$. Evidently, the result
(16) is not, since the boundary condition (14) breaks, in general, the
charge conjugation symmetry. However, for certain choices of the
parameter $\Theta$ this symmetry can be retained [14].

In particular, choosing
\begin{equation}
\left.\begin{array}{ll}
\Theta={\pi\over2}({\rm mod}2\pi),&\Phi^{(0)}-\Upsilon>0\\
\Theta=-{\pi\over2}({\rm mod}2\pi),&\Phi^{(0)}-\Upsilon<0\end{array}\right\}
\quad (\Phi^{(0)}-\Upsilon\neq n,\ \ n\in{{\rm Z}}),
\end{equation}
which corresponds to the boundary condition of
Refs.[15,16], one obtains [12, 13, 17]
\begin{equation}
\langle N\rangle= \left\{\begin{array}{ll} -{1\over2}{\rm
sgn}(m)F,&\Phi^{(0)}-\Upsilon>0\\ {1\over2}{\rm
sgn}(m)(1-F),&\Phi^{(0)}-\Upsilon<0\end{array}\right\},\quad 0<F<1,
\end{equation} which is odd under the charge conjugation but is not periodic in
$\Phi^{(0)}$.

No wonder that there exists a choice of $\Theta$ respecting both the
periodicity in $\Phi^{(0)}$ and the charge conjugation symmetry,
namely,
\begin{equation}\begin{array}{ll}
\Theta={\pi\over2}({\rm mod}2\pi),&0<F<{1\over2}\\
\Theta=-{\pi\over2}[1-{\rm sgn}(m)]({\rm mod}2\pi),&F={1\over2}\\
\Theta=-{\pi\over2}({\rm mod}2\pi),&{1\over2}<F<1\end{array},
\end{equation}
which corresponds to the condition of minimal irregularity, i.e. to
the radial functions being divergent at $r\to0$ at most as $r^{-p}$
with $p\leq{1\over2}$. This is the boundary condition, with the use of
which the result of Ref.[5] is obtained:
\begin{equation}
\langle N\rangle={1\over2}{\rm sgn}(m)\biggl[{1\over2}{\rm
sgn}_0\biggl(F-{1\over2}\biggr)-F+{1\over2}\biggr],
\end{equation}
where
$
{\rm sgn}_0(u)=\left\{
\begin{array}{ll}
{\rm sgn}(u),&u\neq0\\
0,&u=0\end{array}\right. .
$
Note that Eq.(25) is continuous at integer values of
$\Phi^{(0)}-\Upsilon$ and discontinuous at half-integer ones.

Another choice compatible with the periodicity in
$\Phi^{(0)}$ and the symmetry (21) is \linebreak $\Theta=-{\pi\over2}
[1-{\rm sgn}(m)]({\rm mod}2\pi)$ for $0<F<1$; then the vacuum fermion
number is discontinuous both at integer and half-integer values of 
$\Phi^{(0)}-\Upsilon$.

We have calculated also the total magnetic flux induced in the
fermionic vacuum on a punctured plane
\begin{equation}
\Phi^{(I)}=-{e^2F(1-F)\over2\pi|m|}\left[{1\over6}\biggl(F-{1\over2}\biggr)+
{1\over4\pi}\int\limits_1^\infty{dv\over
v\sqrt{v-1}}{Av^F-A^{-1}v^{1-F}\over Av^F+2{\rm
sgn}(m)+A^{-1}v^{1-F}}\right];
\end{equation}
note that the coupling constant $e$ relating the vacuum current to the
vacuum magnetic field strength (via the Maxwell equation) has the
dimension $\sqrt{|m|}$ in $2+1$-dimensional space-time. At
half-integer values of $\Phi^{(0)}-\Upsilon$ we get
\begin{equation}
\Phi^{(I)}=-{e^2\over8\pi^2m}\arctan\biggl\{\tan\biggl[{\Theta\over2}+
{\pi\over4}\bigl(1-{\rm sgn}(m)\bigr)\biggr]\biggr\}.
\end{equation}
The vacuum magnetic flux under the boundary condition (22) is given in
Ref.[12]. Under the boundary condition (24) we get
\begin{equation}
\Phi^{(I)}={e^2F(1-F)\over12\pi|m|}\biggl[{3\over2}{\rm
sgn}_0\bigl(F-{1\over2}\bigr)-F+{1\over2}\biggr],
\end{equation}
which is both periodic in $\Phi^{(0)}$ and $C$-odd.

Thus, we conclude that quantum numbers induced  by a singular magnetic
vortex in the fermionic vacuum depend on the gauge invariant
quantities,  $\Phi^{(0)}-\Upsilon$ and $\Theta$. For certain choices
of $\Theta$ the vacuum quantum numbers are periodic in
$\Phi^{(0)}-\Upsilon$ and have definite $C$-parity.

\bigskip

I would like to thank A.V.~Mishchenko, V.V.~Skalozub and W.I.~Skrypnik
for interesting discussions. The research was partly supported by the
State Committee on Science, Technologies and Industrial Policy of
Ukraine and the American Physical Society.


\begin{thebibliography}{99}
\bibitem{17} It is amusing to note that Eq.(23) has been obtained
by the authors of Ref.[12] at the intermediate stage of
calculations.  However, then these authors add a phantom contribution
to the correct expression and get the incorrect one as a final result;
the latter looks like Eq.(3) which is irrelevant to the case
considered. As to the authors of Ref.[13], who have obtained
also Eq.(23), they, in addition, identify erroneously the limiting
value of this expression at $F \to 1$ with the value of $\langle N\rangle$
at $\Phi^{(0)}-\Upsilon=1$. (Note that the authors of both
Refs.[12,13] consider the case of $\Upsilon=0$).
\end{thebibliography}
\end{document}